\documentclass[
 reprint,
 superscriptaddress,
showpacs,preprintnumbers,
 amsmath,amssymb,
 aps,
 prl,
]{revtex4-1}
\usepackage{bm}
\usepackage[normalem]{ulem} 
\usepackage{color}
\usepackage{dcolumn}
\usepackage{graphicx}
\usepackage{tabularx}
\usepackage{lineno}
\usepackage{hyperref}
\hypersetup{colorlinks=true, citecolor=blue, urlcolor=blue, linkcolor=blue}
\usepackage{booktabs} 
\usepackage{amsmath,amssymb}
\usepackage{comment}

\begin{document}

\title{Precision Measurement of the Beam-Normal Single-Spin Asymmetry in Forward-Angle Elastic Electron-Proton Scattering}

\author{D.~Androi\'c}
\affiliation{University of Zagreb, Zagreb, HR 10002 Croatia } 
\author{D.S.~Armstrong} 
\email{corresponding author: armd@jlab.org}
\affiliation{William \& Mary, Williamsburg, VA 23185 USA}
\author{A.~Asaturyan}
\affiliation{A.~I.~Alikhanyan National Science Laboratory (Yerevan Physics Institute), Yerevan 0036, Armenia}
\author{K. Bartlett}
\affiliation{William \& Mary, Williamsburg, VA 23185 USA}
\author{J.~Beaufait}
\affiliation{Thomas Jefferson National Accelerator Facility, Newport News, VA 23606 USA}
\author{R.S.~Beminiwattha}
\affiliation{Ohio University, Athens, OH 45701 USA}
\affiliation{Louisiana Tech University, Ruston, LA 71272 USA}
\author{J.~Benesch}
\affiliation{Thomas Jefferson National Accelerator Facility, Newport News, VA 23606 USA}
\author{F.~Benmokhtar}
\affiliation{Duquesne University, Pittburgh, PA 15282, USA}
\author{J.~Birchall}
\affiliation{University of Manitoba, Winnipeg, MB R3T2N2 Canada}
\author{R.D.~Carlini}
\affiliation{Thomas Jefferson National Accelerator Facility, Newport News, VA 23606 USA}
\affiliation{William \& Mary, Williamsburg, VA 23185 USA}
\author{J.C.~Cornejo}
\affiliation{William \& Mary, Williamsburg, VA 23185 USA}
\author{S.~Covrig Dusa}
\affiliation{Thomas Jefferson National Accelerator Facility, Newport News, VA 23606 USA}
\author{M.M.~Dalton}
\affiliation{University of Virginia,  Charlottesville, VA 22903 USA}
\affiliation{Thomas Jefferson National Accelerator Facility, Newport News, VA 23606 USA}
\author{C.A.~Davis}
\affiliation{TRIUMF, Vancouver, BC V6T2A3 Canada}
\author{W.~Deconinck}
\affiliation{William \& Mary, Williamsburg, VA 23185 USA}
\author{J.F.~Dowd}
\affiliation{William \& Mary, Williamsburg, VA 23185 USA}
\author{J.A.~Dunne}
\affiliation{Mississippi State University,  Mississippi State, MS 39762  USA}
\author{D.~Dutta}
\affiliation{Mississippi State University,  Mississippi State, MS 39762  USA}
\author{W.S.~Duvall}
\affiliation{Virginia Polytechnic Institute \& State University, Blacksburg, VA 24061 USA}
\author{M.~Elaasar}
\affiliation{Southern University at New Orleans, New Orleans, LA 70126 USA}
\author{W.R.~Falk}
\altaffiliation{deceased}
\affiliation{University of Manitoba, Winnipeg, MB R3T2N2 Canada}
\author{J.M.~Finn}
\altaffiliation{deceased}
\affiliation{William \& Mary, Williamsburg, VA 23185 USA}
\author{T.~Forest}
\affiliation{Idaho State University, Pocatello, ID 83209 USA}
\affiliation{Louisiana Tech University, Ruston, LA 71272 USA}
\author{C. Gal}
\affiliation{University of Virginia,  Charlottesville, VA 22903 USA}
\author{D.~Gaskell}
\affiliation{Thomas Jefferson National Accelerator Facility, Newport News, VA 23606 USA}
\author{M.T.W.~Gericke}
\affiliation{University of Manitoba, Winnipeg, MB R3T2N2 Canada}
\author{J.~Grames}
\affiliation{Thomas Jefferson National Accelerator Facility, Newport News, VA 23606 USA}
\author{V.M.~Gray}
\affiliation{William \& Mary, Williamsburg, VA 23185 USA}
\author{K.~Grimm}
\affiliation{Louisiana Tech University, Ruston, LA 71272 USA}
\affiliation{William \& Mary, Williamsburg, VA 23185 USA}
\author{F.~Guo}
\affiliation{Massachusetts Institute of Technology,  Cambridge, MA 02139 USA}
\author{J.R.~Hoskins}
\affiliation{William \& Mary, Williamsburg, VA 23185 USA}
\author{D.~Jones}
\affiliation{University of Virginia,  Charlottesville, VA 22903 USA}
\author{M.K.~Jones}
\affiliation{Thomas Jefferson National Accelerator Facility, Newport News, VA 23606 USA}
\author{R.T.~Jones}
\affiliation{University of Connecticut,  Storrs-Mansfield, CT 06269 USA}
\author{M.~Kargiantoulakis}
\affiliation{University of Virginia,  Charlottesville, VA 22903 USA}
\author{P.M.~King}
\affiliation{Ohio University, Athens, OH 45701 USA}
\author{E.~Korkmaz}
\affiliation{University of Northern British Columbia, Prince George, BC V2N4Z9 Canada}
\author{S.~Kowalski}
\affiliation{Massachusetts Institute of Technology,  Cambridge, MA 02139 USA}
\author{J.~Leacock}
\affiliation{Virginia Polytechnic Institute \& State University, Blacksburg, VA 24061 USA}
\author{J.P.~Leckey}
\affiliation{William \& Mary, Williamsburg, VA 23185 USA}
\author{A.R.~Lee}
\affiliation{Virginia Polytechnic Institute \& State University, Blacksburg, VA 24061 USA}
\author{J.H.~Lee}
\affiliation{Ohio University, Athens, OH 45701 USA}
\affiliation{William \& Mary, Williamsburg, VA 23185 USA}
\author{L.~Lee}
\affiliation{TRIUMF, Vancouver, BC V6T2A3 Canada}
\affiliation{University of Manitoba, Winnipeg, MB R3T2N2 Canada}
\author{S.~MacEwan}
\affiliation{University of Manitoba, Winnipeg, MB R3T2N2 Canada}
\author{D.~Mack}
\affiliation{Thomas Jefferson National Accelerator Facility, Newport News, VA 23606 USA}
\author{J.A.~Magee}
\affiliation{William \& Mary, Williamsburg, VA 23185 USA}
\author{R.~Mahurin}
\affiliation{University of Manitoba, Winnipeg, MB R3T2N2 Canada}
\author{J.~Mammei}
\affiliation{University of Manitoba, Winnipeg, MB R3T2N2 Canada}
\affiliation{Virginia Polytechnic Institute \& State University, Blacksburg, VA 24061 USA}
\author{J.W.~Martin}
\affiliation{University of Winnipeg, Winnipeg, MB R3B2E9 Canada}
\author{M.J.~McHugh}
\affiliation{George Washington University, Washington, DC 20052 USA}
\author{D.~Meekins}
\affiliation{Thomas Jefferson National Accelerator Facility, Newport News, VA 23606 USA}
\author{J.~Mei}
\affiliation{Thomas Jefferson National Accelerator Facility, Newport News, VA 23606 USA}
\author{K.E.~Mesick}
\affiliation{George Washington University, Washington, DC 20052 USA}
\affiliation{Rutgers, the State University of New Jersey, Piscataway, NJ 088754 USA}
\author{R.~Michaels}
\affiliation{Thomas Jefferson National Accelerator Facility, Newport News, VA 23606 USA}
\author{A.~Micherdzinska}
\affiliation{George Washington University, Washington, DC 20052 USA}
\author{A.~Mkrtchyan}
\affiliation{A.~I.~Alikhanyan National Science Laboratory (Yerevan Physics Institute), Yerevan 0036, Armenia}
\author{H.~Mkrtchyan}
\affiliation{A.~I.~Alikhanyan National Science Laboratory (Yerevan Physics Institute),
Yerevan 0036, Armenia}
\author{N.~Morgan}
\affiliation{Virginia Polytechnic Institute \& State University, Blacksburg, VA 24061 USA}
\author{A.~Narayan}
\affiliation{Mississippi State University,  Mississippi State, MS 39762  USA}
\author{L.Z.~Ndukum}
\affiliation{Mississippi State University,  Mississippi State, MS 39762  USA}
\author{V.~Nelyubin}
\affiliation{University of Virginia,  Charlottesville, VA 22903 USA}
\author{Nuruzzaman}
\affiliation{Hampton University, Hampton, VA 23668 USA}
\affiliation{Mississippi State University,  Mississippi State, MS 39762  USA}
\author{W.T.H van Oers}
\affiliation{TRIUMF, Vancouver, BC V6T2A3 Canada}
\affiliation{University of Manitoba, Winnipeg, MB R3T2N2 Canada}
\author{V.F. Owen} 
\affiliation{William \& Mary, Williamsburg, VA 23185 USA}\author{S.A.~Page}
\affiliation{University of Manitoba, Winnipeg, MB R3T2N2 Canada}
\author{J.~Pan}
\affiliation{University of Manitoba, Winnipeg, MB R3T2N2 Canada}
\author{K.D.~Paschke}
\affiliation{University of Virginia,  Charlottesville, VA 22903 USA}
\author{S.K.~Phillips}
\affiliation{University of New Hampshire, Durham, NH 03824 USA}
\author{M.L.~Pitt}
\affiliation{Virginia Polytechnic Institute \& State University, Blacksburg, VA 24061 USA}
\author{R.W. Radloff}
\affiliation{Ohio University, Athens, OH 45701 USA}
\author{J.F.~Rajotte}
\affiliation{Massachusetts Institute of Technology,  Cambridge, MA 02139 USA}
\author{W.D.~Ramsay}
\affiliation{TRIUMF, Vancouver, BC V6T2A3 Canada}
\affiliation{University of Manitoba, Winnipeg, MB R3T2N2 Canada}
\author{J.~Roche}
\affiliation{Ohio University, Athens, OH 45701 USA}
\author{B.~Sawatzky}
\affiliation{Thomas Jefferson National Accelerator Facility, Newport News, VA 23606 USA}
\author{T.~Seva}
\affiliation{University of Zagreb, Zagreb, HR 10002 Croatia } 
\author{M.H.~Shabestari}
\affiliation{Mississippi State University,  Mississippi State, MS 39762  USA}
\author{R.~Silwal}
\affiliation{University of Virginia,  Charlottesville, VA 22903 USA}
\author{N.~Simicevic}
\affiliation{Louisiana Tech University, Ruston, LA 71272 USA}
\author{G.R.~Smith}
\affiliation{Thomas Jefferson National Accelerator Facility, Newport News, VA 23606 USA}
\author{P.~Solvignon}
\altaffiliation{deceased}
\affiliation{Thomas Jefferson National Accelerator Facility, Newport News, VA 23606 USA}
\author{D.T.~Spayde}
\affiliation{Hendrix College, Conway, AR 72032 USA}
\author{A.~Subedi}
\affiliation{Mississippi State University,  Mississippi State, MS 39762  USA}
\author{R.~Subedi}
\affiliation{George Washington University, Washington, DC 20052 USA}
\author{R.~Suleiman}
\affiliation{Thomas Jefferson National Accelerator Facility, Newport News, VA 23606 USA}
\author{V.~Tadevosyan}
\affiliation{A.~I.~Alikhanyan National Science Laboratory (Yerevan Physics Institute),
Yerevan 0036, Armenia}
\author{W.A.~Tobias}
\affiliation{University of Virginia,  Charlottesville, VA 22903 USA}
\author{V.~Tvaskis}
\affiliation{University of Winnipeg, Winnipeg, MB R3B2E9 Canada}
\author{B.~Waidyawansa}
\affiliation{Ohio University, Athens, OH 45701 USA}
\affiliation{Louisiana Tech University, Ruston, LA 71272 USA}
\author{P.~Wang}
\affiliation{University of Manitoba, Winnipeg, MB R3T2N2 Canada}
\author{S.P.~Wells}
\affiliation{Louisiana Tech University, Ruston, LA 71272 USA}
\author{S.A.~Wood}
\affiliation{Thomas Jefferson National Accelerator Facility, Newport News, VA 23606 USA}
\author{S.~Yang}
\affiliation{William \& Mary, Williamsburg, VA 23185 USA}
\author{P. Zang}
\affiliation{Syracuse University, Syracuse, NY 13244
USA}
\author{S.~Zhamkochyan}
\affiliation{A.~I.~Alikhanyan National Science Laboratory (Yerevan Physics Institute), Yerevan 0036, Armenia}

\collaboration{The Q$_{\rm weak}$ Collaboration}
\noaffiliation

\date{\today}

\begin{abstract}
A beam-normal single-spin asymmetry generated in the scattering of transversely polarized electrons from unpolarized nucleons is an observable related to the imaginary part of the two-photon exchange process. 
We report a 2\% 
precision 
measurement of the beam-normal single-spin asymmetry in elastic electron-proton scattering with a mean scattering angle of 
$\theta_{\rm lab} = 7.9^{\circ}$ and a mean energy of 1.149 GeV. The asymmetry result is $B_n =  -5.194 \pm 0.067 \; \rm{ (stat)}  \pm 0.082$ (syst) ppm.
This is the most precise measurement of this quantity available to date and therefore provides a stringent test of two-photon exchange models at far-forward scattering angles ($\theta_{\rm lab} \rightarrow 0$)
where they should be most reliable. 

\end{abstract}

\maketitle

The high intensities of electron beams at facilities like Jefferson Lab and MAMI make them ideal for studying the charge and magnetization distributions inside nuclear matter in the single-photon exchange (Born) approximation. However, high precision measurements can be affected by two-photon exchange (TPE)
\cite{Afanasev:2017gsk}.
Depending on the observable, either the real or imaginary part of the TPE amplitude can play a role.

There has been significant effort to study the real part of the TPE amplitude because it affects cross sections \cite{Afanasev:2017gsk}. However, the uncertainties in the theoretical calculations are large, and constraints on models remain weak even after a decade-long program of targeted measurements \cite{Afanasev:2017gsk}. 
An alternative approach is to study observables 
proportional to the imaginary part of the TPE amplitude such as the beam-normal single-spin asymmetry (BNSSA, $A_y$~\cite{osti_4726823}, or just $B_n$) .

$B_n$ is a parity- and CP-conserving  asymmetry typically at the few part-per-million (ppm) level 
for forward angles and GeV-scale incident energies in $\vec{e}p$ elastic scattering. Required by time-reversal invariance to vanish in the one-photon exchange approximation, a non-zero $B_n$ can only arise with the exchange of two or more photons between the scattered electron and the target nucleon \cite{Gorchtein:2004ac}.  
Experimentally, $B_n$ manifests itself as the amplitude of an azimuthal variation of the asymmetry 
when the beam is polarized transverse to its incident momentum. 

Theoretically,  two complementary approaches have been pursued. One \cite{Pasquini:2004pv,priv:Pasquni2013} is expected to be valid at all angles, but should work best at lower energies because it only includes the $\pi$N intermediate state as well as the (smaller) elastic proton contribution. The other approach 
\cite{Gorchtein:2004ac,Gorchtein:2005yz,Gorchtein:2014hla,GORCHTEIN2004234,GorchPC,Afanasev:2017gsk,Afanasev:2004pu,AfanasevPC} is expected to work at all energies because it includes contributions from multi-particle intermediate states (e.g. $\pi \pi$N, $\eta$N, K$\Lambda$, ...), 
but works best at forward angles because it uses the optical theorem to relate the measured total photoproduction cross-section to the imaginary part of the TPE forward scattering amplitude $\mathcal{I}m$(TPE). 
    
Hard TPE was generally treated as causing small (percent-level) corrections to the unpolarized scattering cross-section that are independent of hadronic structure \cite{Mo:1968cg,Maximon:2000hm}.
However in 2000 a striking disagreement in the proton's elastic electromagnetic form-factor ratio ($G_E^p/G_M^p$) was observed when comparing Rosenbluth (L/T) separation \cite{Rosenbluth:1950yq} and polarization transfer \cite{Jones:1999rz} results at $Q^2\geq$ 2 (GeV/c)$^2$.
 This discrepancy (known as the proton form-factor puzzle) could be explained \cite{Guichon:2003qm} by a correction involving the real part of the TPE amplitude that modifies the Rosenbluth cross-section, but largely cancels in the polarization-transfer ratios. A recent summary can be found in \cite{Afanasev:2017gsk}.
    
The real part of the TPE amplitude $\mathcal{R}e$(TPE) can be determined from the ratio of $e^\pm p$ cross sections (see VEPP-3 \cite{Rachek:2014fam}, OLYMPUS \cite{Henderson:2016dea}, CLAS \cite{Adikaram:2014ykv}). In principle $\mathcal{R}e$(TPE) can also be determined from the imaginary part via dispersion relations. In practice this is difficult since a broad range of kinematics is needed and there is a paucity of $B_n$ results. Nevertheless, the effects of TPE on the proton radius puzzle (see \cite{Xiong:2019umf} for the most recent results and a summary)
have been explored theoretically \cite{Gorchtein:2014hla} using an unsubtracted fixed-$t$ dispersion relation to do just that, predicting that TPE effects are at the level of the present uncertainties ($\approx 1\%$) in the proton radius determinations from $ep$ scattering data. Future experiments (MUSE \cite{Gilman:2013eiv,Gilman:2017hdr}) aim to improve this precision and further explore TPE effects by comparing $e^\pm p$ and $\mu^\pm p$ scattering.
This underscores the importance of providing $B_n$ data to test the predictions of $\mathcal{I}m$(TPE).

The kinematics of this experiment are at a far-forward electron scattering angle ($7.9^\circ$) where the optical model approach should work well, and with a small four-momentum transfer $Q^2=-t=0.0248$ (GeV/c)$^2$, and an intermediate energy ($E_{\rm lab}=1.149$ GeV,  $E_{\rm cm}=1.74$ GeV) where up to five pion intermediate states can contribute.
The asymmetry is generated by the interference of  one-photon and two-photon exchange processes and has the form \cite{DeRujula:1972te}

\vspace*{-0.3cm}
\begin{equation}
\label{eq:normal_spin_formula}
B_n = \frac{\sigma^\uparrow -  \; \sigma^\downarrow}{\sigma^\uparrow +  \; \sigma^\downarrow} = \frac{2 \; \mathcal{I}m ( \mathcal{M}_{\gamma \gamma} \mathcal{M}_{\gamma}^* )}{|\mathcal{M}_{\gamma}| ^2},
\end{equation}
\noindent
where 
$\sigma^\uparrow (\sigma^\downarrow)$ denotes the scattering cross section for electrons with spin parallel (anti-parallel) to a vector $\hat{n}$ normal to the scattering plane, where $\hat{n}=(\vec{k}\times\vec{k^{\prime}})/(|\vec{k}\times\vec{k^{\prime}}|)$ with $\vec{k}(\vec{k^{\prime}})$ being the momentum of the incoming(outgoing) electron.  $\mathcal{M}_{\gamma}$ and $\mathcal{M}_{\gamma \gamma}$ are the amplitudes for one- and two-photon exchange.  
For transversely-polarized electrons scattering from unpolarized nucleons, the detected asymmetry then depends on the azimuthal scattering angle $\phi$ via $A_{\rm exp}(\phi) \approx B_n \vec{P}\cdot\hat{n}$,
where $\vec{P}$ is the electron polarization vector.

Companion measurements of $B_n$ are necessary in most parity-violating electron scattering experiments in order to account for the effects of residual transverse polarization in the nominally longitudinally-polarized beam.
Previous measurements of $B_n$ at far-forward angles ($6.0^\circ\! < \!\theta_{\rm lab} \!<\! 9.7^\circ$) were obtained by the G0~\cite{Armstrong:2007vm} and HAPPEX~\cite{Abrahamyan:2012cg} collaborations with E$_{\rm lab} $ near 3 GeV. Somewhat larger-angle results have been obtained by  PVA4~\cite{Maas:2004pd,Gou:2020viq} for $(\theta_{\rm lab}$, E$_{\rm lab})=(\approx\!34^\circ,0.3-1.5$ GeV), 
and by SAMPLE~\cite{Wells:2000rx} at ($\approx\! 55^\circ, 0.2$ GeV). Backward angle experiments were performed at 
$(108^\circ,0.36$ and 0.69 GeV) by G0~\cite{Androic:2011rh}, and at $(145^\circ,0.32$ and  0.42 GeV) by PVA4~\cite{Rios:2017vsw}. Some of these experiments also included results on deuterium~\cite{Androic:2011rh,Rios:2017vsw} as well as heavier nuclei~\cite{Abrahamyan:2012cg}.

The ($7.9^\circ, 1.149$ GeV) elastic $\vec{e}p$ $B_n$ measurement reported here was part of a series of 
companion measurements performed by the Q$_{\rm weak}$ collaboration to constrain systematic uncertainties in the first determination of the weak charge of the proton \cite{Qweak.Nature,Androic:2013rhu}. The general performance of the experimental apparatus is described in Ref.~\cite{ALLISON2015105}. Details relevant to the extraction of $B_n$ are presented here.

A total of 54 hours of $B_n$ data were collected in three measurement periods and with two different orientations of  transverse polarization.
Polarized electrons were generated by photo-emission from a strained GaAs cathode at the injector of the Thomas Jefferson National Accelerator Facility. Two Wien filters \cite{Grames:2011} were used to rotate the electron spin 
in the transverse plane to horizontal (spin pointing to beam-right at the target) or vertical (spin pointing up). The transversely polarized, 150 $\mu A$ - 180 $\mu A$ electron beam was then accelerated to 
1.16 GeV before reaching the Q$_{\rm weak}$ apparatus in experimental Hall C. There it scattered
from unpolarized liquid hydrogen encased in a 34.4-cm-long aluminum-alloy cell with thin (0.1 mm thick) windows where the beam entered and exited. Longitudinal polarization measurements  (bracketing the transverse running)
using  $\mathrm{M\o ller}$  and Compton polarimeters~\cite{Hauger:1999iv,Magee:2016xqx,Narayan:2015aua}
upstream of the target yielded an average  statistics-weighted beam polarization 
$\langle$P$\rangle$ = 
(88.72 $\pm$ 0.70)\%. 
During the transverse running, the polarization was verified to be $>99.97\%$ transverse via null measurements with the $\mathrm{M\o ller}$  polarimeter, which is only sensitive to longitudinal beam polarization.

A set of collimators located downstream of the target selected electrons with lab scattering angles of 5.8$^{\circ}$ to 11.6$^{\circ}$. 
A toroidal magnet then focused  elastic electrons onto a set of eight Cherenkov detectors placed symmetrically around the beam axis, 12.2 m downstream of the target. The azimuthal coverage of the detector array was 49\% of 2$\pi$.

 The spin direction of the electrons was 
selected  from one of two pseudo-randomly chosen quartet patterns ($\uparrow \downarrow \downarrow \uparrow$ or $\downarrow \uparrow \uparrow \downarrow$)
generated at 240 Hz. Here 
$\uparrow$ represents the standard spin orientation (spin up or to beam right) and $\downarrow$ represents a 180$^{\circ}$ rotation in the corresponding plane.
The signals from the Cherenkov detectors were integrated for each $\uparrow$ and $\downarrow$ spin state  (at 960 Hz). The detector asymmetries were calculated for each quartet using $A_{\rm raw} = \frac{Y_\uparrow - \,Y_\downarrow}{Y_\uparrow + \,Y_\downarrow}$
where $Y_{\uparrow (\downarrow)}$ is the charge-normalized detector yield in the $\uparrow$ or $\downarrow$ spin state.
The systematic uncertainty due to the beam charge normalization was negligible here~\cite{Qweak.Nature}.
False asymmetries from spin-correlated beam position, angle, and energy changes were largely cancelled by the periodic insertion of a half-wave plate (IHWP) located in the injector. Further suppression of false asymmetries was achieved by using $A_{\rm msr} = A_{\rm raw} - \sum\limits^5_{j=1} \left( \frac{\partial A}{\partial \chi_j} \right) \Delta \chi_j$ where $\Delta \chi_j$ are the helicity-correlated differences in  
beam position (vertical and horizontal), beam angle (vertical and horizontal) and beam energy
over the helicity quartet, and the slopes $\partial A / \partial \chi_j$ were 
determined using multi-variable linear regression \cite{BWaidyawansa_phd}. False asymmetries caused by secondary events scattered from beamline
elements were negligible ($< 0.005 {\rm \: ppm})$
~\cite{Qweak.Nature}.

The measured asymmetries $A^i_{\rm msr}$ in detector $i$, for both orientations of the transverse beam polarization,  were fit to
\begin{equation}
\label{eq:bnsa_phi_dependance}
A^i_{\rm msr}(\phi_i) = 
R_{l} R_{\rm av} A_{\rm exp}\sin(\phi_{\rm s}-\phi_i + \phi_{\rm off}) + C,
\end{equation} 
to extract the experimental asymmetry $A_{\rm exp}$.
Here 
 $\phi_{s}$ is the azimuthal angle of $\vec{P}$, and $\phi_i$ is the azimuthal angle of the $i^{th}$ detector in the plane normal to the beam axis. The factor $R_{\rm av} = 0.9938 \pm 0.0006$ accounts for the averaging of the asymmetry over the effective azimuthal acceptance ($\approx$ 22$^{\circ}$) of a  Cherenkov detector and $R_l = 1.007 \pm 0.005$ corrects for the
measured non-linearity in the detector electronics. A floating offset in phase $\phi_{\rm off}$ was included to account for any detector offsets in the azimuthal plane, and a floating constant $C$ was included to represent any monopole asymmetries, such as due to parity-violating asymmetry generated by any residual longitudinal beam polarization. The fitted values for $\phi_{\rm off}$ and $C$ were consistent with zero, and the
value of $A_{\rm exp}$ extracted was  
insensitive to the inclusion of these extra fit parameters. 

\begin{figure}[b]
\centering
\includegraphics[width=1.0\columnwidth]{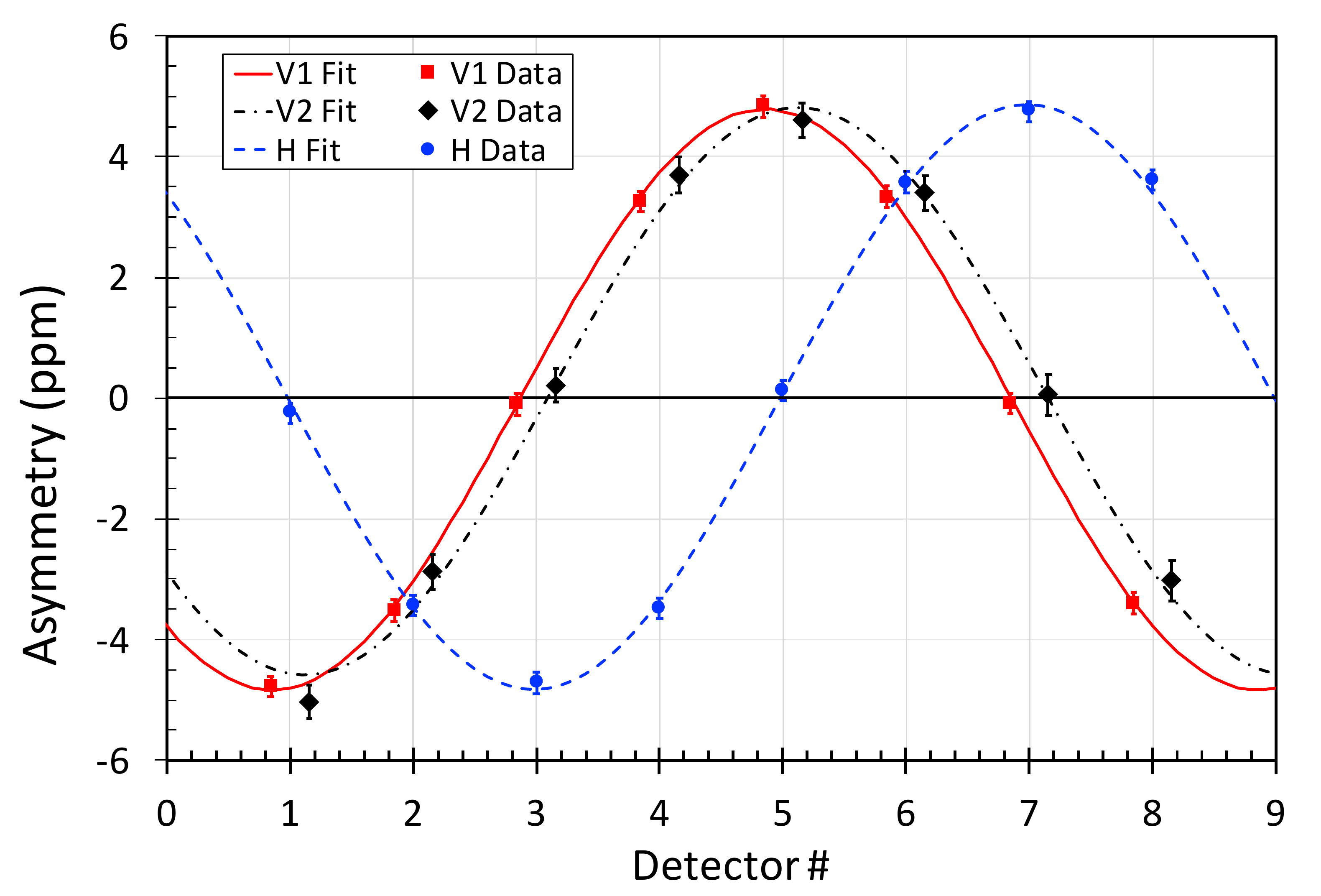}\caption{\label{fig:compare_v_h}Extraction of the experimental asymmetry $A_{\rm exp}$ from the measured asymmetries $A^i_{\rm msr}$ for the horizontal (H) and  the two vertical data sets (V1 and V2). The phases of the vertical data sets were offset $-7^\circ$ (V1) and $+7^\circ$ (V2) in the figure for clarity.
The detector number corresponds to the azimuthal location of the detectors, starting from beam left (Detector 1) where $\phi_i=0^\circ$, and increasing clockwise every $45^\circ$. Uncertainties shown are statistical only. The reduced $\chi^2$ (5 degrees of freedom) in the 
fits are 0.15 (V1), 1.07 (V2), and 0.81 (H). }
\end{figure}

The fits to Eq.~\ref{eq:bnsa_phi_dependance} for
the three data sets are shown in Fig.~\ref{fig:compare_v_h}. 
 Since the kinematics were similar and the results consistent, the error-weighted average of the three  measurements
$A_{\rm exp} = -4.801 \pm 0.056$ (stat) $\pm$ 0.039 (syst) ppm 
was used as the experimental asymmetry from the full measurement. The systematic error accounts for the uncertainties in $R_l$, $R_{\rm av}$ and the linear regression.

The experimental asymmetry $A_{\rm exp}$ was then corrected for four backgrounds.  The largest background 
was $f_1= 3.3\pm$0.2\%, a dilution from elastic and quasi-elastic electrons scattering from the aluminum-alloy beam-entrance and exit windows of the 
target. Dedicated measurements using an 
aluminum-alloy target, similar to but thicker than the windows used in the target cell, were used to determine the aluminum asymmetry $A_1$ \cite{BWaidyawansa_phd}.
Another background correction was applied for $f_2 = 0.018 \pm 0.004$\%, a dilution due to  inelastic electrons. The inelastic asymmetry $A_2$ \cite{BWaidyawansa_phd} was determined using dedicated measurements with the toroidal magnet configured to focus inelastic electrons onto the detectors. Additionally, neutral backgrounds in the acceptance generated by sources in the beamline ($f_3 = 0.19 \pm 0.06$\% dilution) and other sources ($f_4 < 0.3$\% dilution) were studied.
These neutral backgrounds constituted negligible corrections to the experiment's final  azimuthal asymmetry. Therefore, no correction was applied ($A_3 \approx A_4 \approx 0$). However, their dilutions were taken into consideration. 

A unique potential background asymmetry not yet observed in a $B_n$ measurement is a parity-violating 
beam-transverse single-spin asymmetry ($A_x$), generated by the interference between one-photon exchange and the Z$^0$-exchange processes. 
 At our kinematics, $A_x$ is estimated to be on the order of $10^{-11}$ \cite{priv:WBP},
too small to be observed in this experiment. 

The various corrections were applied to the experimental asymmetry $A_{\rm exp}$ to extract $B_n$ following \begin{equation}
\label{eq:asymmetry_correction_formula}
B_n = R_{\rm tot}\left[\frac{A_{\rm exp}/P - \sum_{i=1}^4 f_iA_i}{1-\sum_{i=1}^4 f_i}\right] + A_{\rm bias}.
\end{equation}
Here $A_i$ is the background asymmetry generated by the $i^{th}$ background (aluminum  windows, inelastics, beamline neutrals, and other  neutrals, respectively) with dilution $f_i$. 
The factor $R_{\rm tot} = 1.0041 \pm 0.0046$ accounts for electron energy-loss and depolarization from electromagnetic radiation, non-uniform $Q^2$ distribution across the detectors, light-collection variation across the detectors, and the uncertainty in the acceptance-averaged $\langle Q^2 
\rangle$ = 0.0248 $\pm$ 0.0001 GeV$^2$.
$A_{\rm bias} = 0.125 \pm 0.041$~ppm is a false asymmetry that arose due to the analyzing power of the scattered electrons that can rescatter in the lead pre-radiators 
installed upstream of each main detector. This effect is described in detail elsewhere~\cite{Qweak.Nature}; it was larger in magnitude in the present case because, for transversely polarized beam, it does not largely cancel due to the symmetry of the apparatus. With the above corrections, we obtain a value of $B_n =  -5.194 \pm 0.067$ (stat) $\pm$ 0.082 (syst) ppm for elastic electron-proton scattering at a vertex scattering angle of $\langle 
\theta \rangle$ = 7.9$^{\circ}$ and vertex energy $\langle E \rangle$ = 1.149 GeV. The contributions from different error sources
are summarized in Table~\ref{tab:bnsa_error_summary} and  discussed in more detail in Ref.~\cite{BWaidyawansa_phd}. 

\begin{table}[!hhhbtb]
\centering
\caption{\label{tab:bnsa_error_summary}Summary of experimental uncertainties.}
\begin{tabular}{lc}
\hline
Uncertainty Source & $\frac{\Delta B_n}{B_n}$ (\%) \\
\hline
Statistics                                    & 1.29 \\
\hline
\multicolumn{2}{c}{Systematics}\\
\hline
$P$: Beam polarization                             & 0.807 \\ \addlinespace[-1.1ex]
$R_{\rm tot}$: Kinematics and acceptance                        & 0.428 \\ \addlinespace[-1.1ex]

$R_l$: Electronic non-linearity                                 & 0.540 \\ \addlinespace[-1.1ex]
Linear regression                             & 0.656 \\ \addlinespace[-1.1ex] 
$R_{\rm av}$: Acceptance averaging & 0.067 \\ \addlinespace[-1.1ex]
$A_1$: Aluminum background asymmetry & 0.408 \\ \addlinespace[-1.1ex]
$f_1$: Aluminum dilution                             & 0.172 \\ \addlinespace[-1.1ex]
$A_2$: Inelastic background asymmetry                & 0.024 \\  \addlinespace[-1.1ex] 
$f_2$: Inelastic dilution                            & 0.030 \\  \addlinespace[-1.1ex]
$A_3$: Beamline neutral asymmetry                 & 0.004 \\ \addlinespace[-1.1ex]
$f_3$: Beamline neutral dilution                  & 0.064 \\ \addlinespace[-1.1ex]
$A_4$: Other neutral background asymmetry             & 0.201 \\ \addlinespace[-1.1ex]
$f_4$: Other neutral background dilution              & 0.213 \\ \addlinespace[-1.1ex]
$A_{\rm bias}$                                & 0.789
\\
\hline
Systematics Sub Total                         & 1.57 \\
\hline
\hline
Total Uncertainty                             & 2.03 \\
\hline
\end{tabular}
\end{table}

\begin{figure}[!t]
\includegraphics[width=1.0\columnwidth]{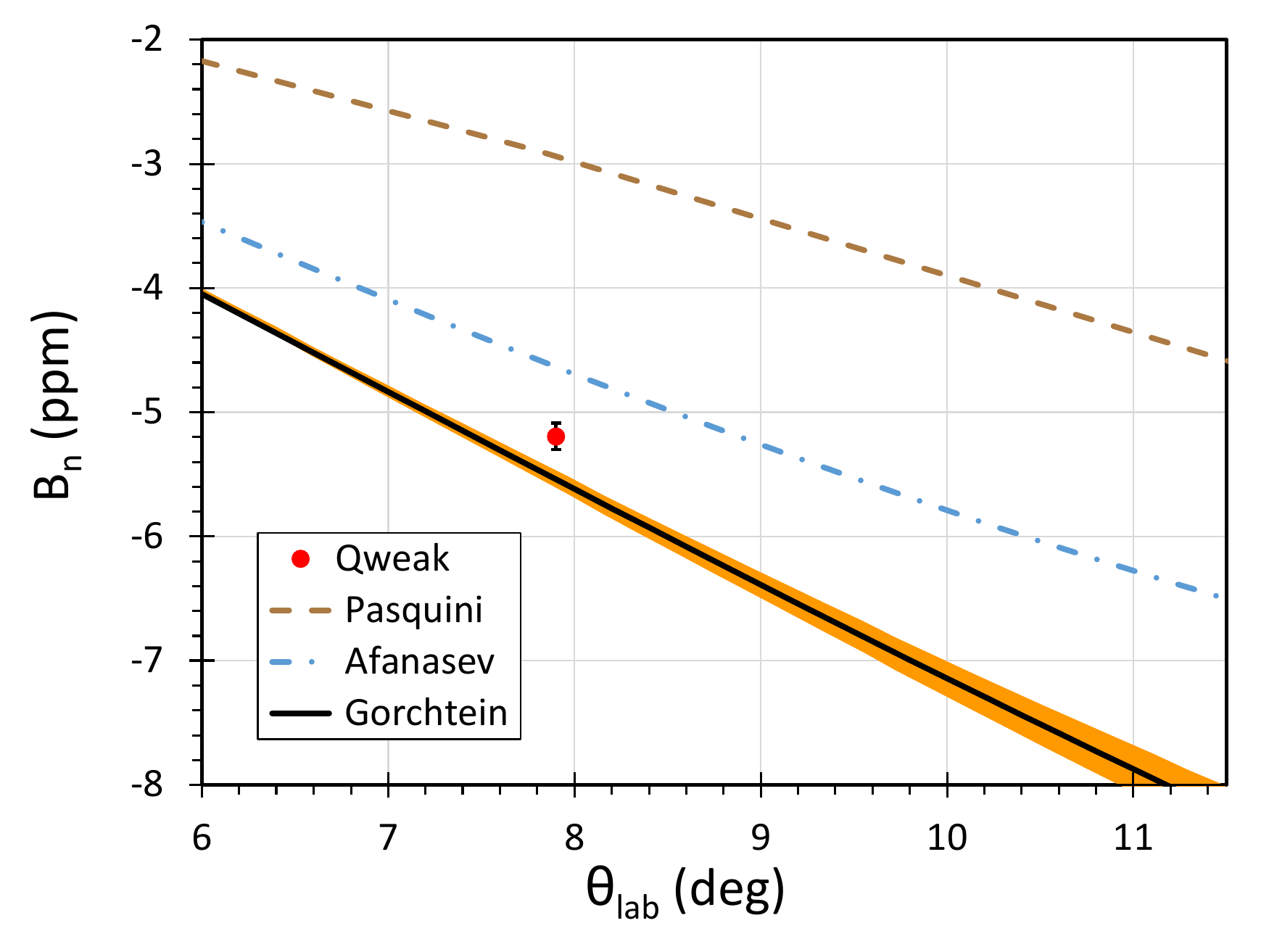}
\caption{\label{fig:trans_compare_theory}Comparison of this measurement (red circle) to calculations at E$_{\rm lab}$=1.149 GeV by  Pasquini \& Vanderhaeghen\  
\cite{Pasquini:2004pv}, Afanasev \& Merenkov\ 
\cite{Afanasev:2004pu}, and Gorchtein\ \cite{Gorchtein:2005yz} over the Q$_{\rm weak}$ acceptance.  The orange band about the latter calculation indicates the model uncertainty. 
}
\end{figure}

Figure~\ref{fig:trans_compare_theory} compares our measurement to three model calculations:  Pasquini \& Vanderhaeghen\ \cite{Pasquini:2004pv,priv:Pasquni2013}, Afanasev \& Merenkov\ \cite{Afanasev:2004pu,AfanasevPC} and Gorchtein\ \cite{Gorchtein:2004ac,Gorchtein:2005yz,Gorchtein:2014hla,GORCHTEIN2004234,GorchPC}. The latter model~\cite{Gorchtein:2004ac,Gorchtein:2005yz,Gorchtein:2014hla,GORCHTEIN2004234,GorchPC} is in closest agreement with this measurement (within 0.3 ppm, or just 7\%), but still 2.7 $\sigma$  away, given the small Q$_{\rm weak}$ uncertainty. 
The other prediction that also uses the optical theorem \cite{Afanasev:2004pu,AfanasevPC} 
is only slightly further away.
The  Pasquini \& Vanderhaeghen\ model significantly underpredicts the magnitude of $B_n$. The latter calculation uses unitarity to model the Doubly Virtual Compton Scattering (VVCS) tensor in the resonance regime in terms of electroabsorption amplitudes whereas both Afanasev \& Merenkov\ as well as
Gorchtein\ use the optical theorem to relate the forward VVCS tensor to the total photoabsorption cross section. Although the three calculations predict similar angular behavior for the asymmetry in our acceptance, their magnitudes vary widely. 

Generally, the models agree that the dominant contribution to the asymmetry comes from the inelastic intermediate states of the nucleon in TPE. The contribution from the elastic state is insignificant. However, both the Afanasev \& Merenkov\ model and the Gorchtein\ model consider all inelastic intermediate states with multi-pion excitations whereas the  Pasquini \& Vanderhaeghen\ model only considers inelastic states with single-pion excitations. This 
likely causes the largest difference between the two types of 
calculations \cite{priv:Pasquni2013,GorchPC,AfanasevPC}.

\begin{figure}[!ttthb]
\includegraphics[width=1.0\columnwidth]{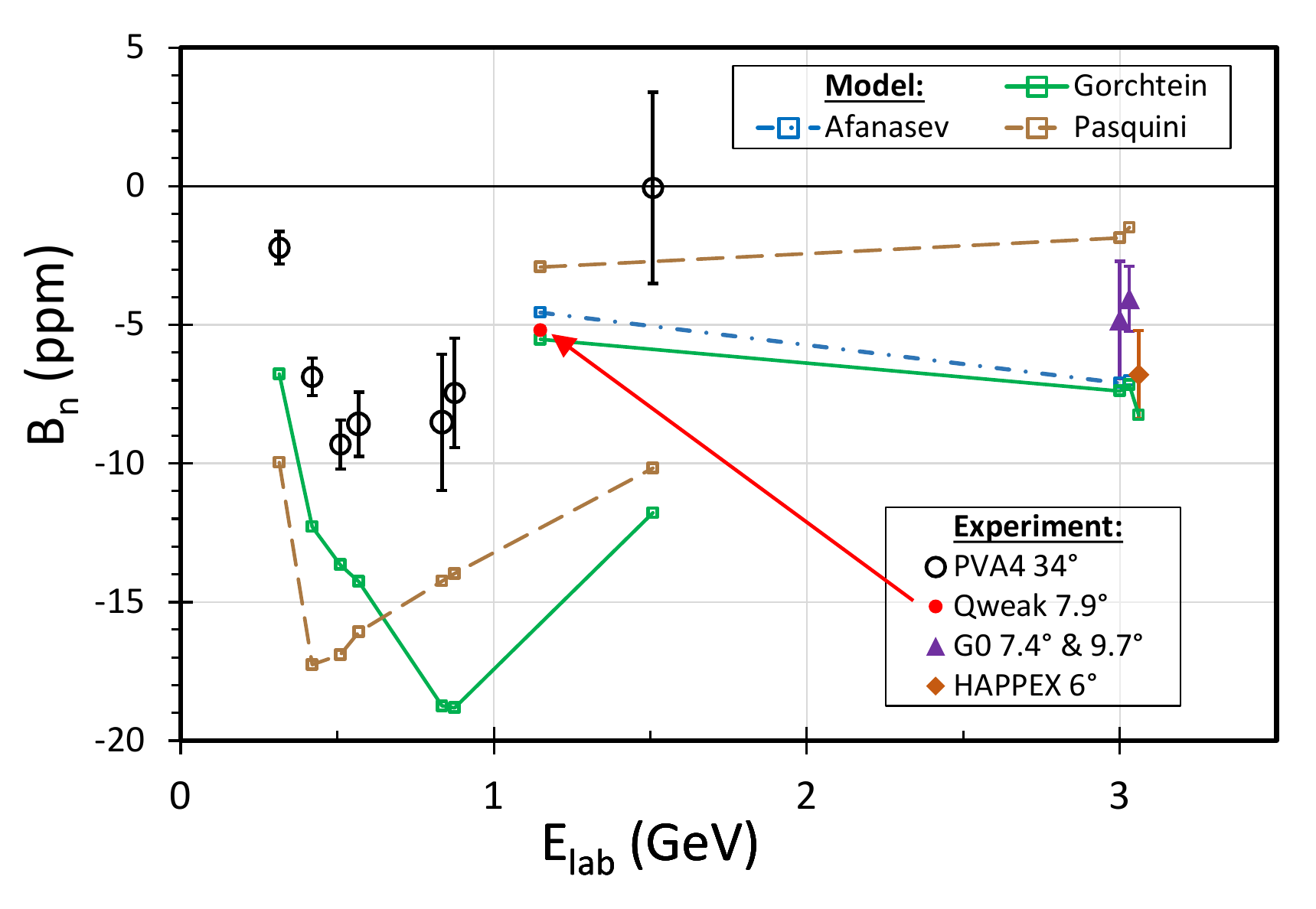}
\caption{
Beam energy dependence of all forward-angle ($\theta_{\rm lab}\le34^\circ$) elastic $\vec{e}p$ $B_n$ data compared to calculations at each experiment's kinematics.  The far-forward angle data (solid symbols, $\theta_{\rm lab}<10^\circ$) are from this experiment (red circle, uncertainty smaller than the symbol), G0~\cite{Armstrong:2007vm} (purple triangles), and HAPPEX~\cite{Abrahamyan:2012cg} (orange diamond). Less forward-angle data $\theta_{\rm lab}\approx34^\circ$ are denoted with open circles from 
PVA4~\cite{Gou:2020viq,Maas:2004pd} (black).
The predictions (open squares) from each  theoretical group (Pasquini \& Vanderhaeghen\ ~\cite{Pasquini:2004pv,priv:Pasquni2013}, Afanasev \& Merenkov\ ~\cite{Afanasev:2004pu,AfanasevPC}, and Gorchtein\ ~\cite{Gorchtein:2004ac,Gorchtein:2005yz,Gorchtein:2014hla,GORCHTEIN2004234,GorchPC}) are connected by straight-line segments 
for the far-forward and the forward angle calculations, 
to help guide the eye. 
 }
 \label{fig:BnvsE}
\end{figure}

The calculations from the three theoretical groups discussed here differ at different kinematics, making a global comparison to other experiments difficult. For example, the Gorchtein model includes corrections to account for the off-forward  $34^\circ$ data of \cite{Gou:2020viq}, which 
are not used to predict the far-forward $7.9^\circ$ kinematics of this experiment.
However, it is still instructive to compare the existing forward angle $B_n$ data to the kinematics-specific predictions from each theoretical group. Such a comparison is shown as a function of $E_{\rm lab}$ in Figure~\ref{fig:BnvsE} for $\theta_{\rm lab} \leq 34^\circ$ data. 
This figure shows that all the models have significant disagreements with the less-forward angle ($\theta_{\rm lab} > 10^\circ$) data.  The far-forward data are in a better position to be described theoretically using the optical theorem and those calculations do show reasonable agreement.  The Q$_{\rm weak}$ result provides by far the most precise test of models to date in the kinematic region where they are expected to be most accurate.

The  beam-normal single-spin asymmetry is a unique tool to test dispersion relations 
used in calculating TPE corrections to $ep$ scattering cross sections. 
In light of improving these TPE corrections in $ep$ and $\mu p$ scattering observables, precision measurements of $B_n$ 
are extremely useful for validating TPE models.
The precise Q$_{\rm weak}$ datum reported here, in particular,  provides a stringent test of the TPE models at far-forward angles and moderate energy. 

We thank the staff of Jefferson Lab, TRIUMF, and MIT Bates, as well as our undergraduate students, for their vital support during this challenging experiment. We also thank B. Pasquini, A. Afanasev, M. Gorchtein, M. Vanderhaeghen, O. Tomalak, P. Blunden and W. Melnitchouk for 
useful discussions. This work was supported by the U.S. Department of Energy (DOE), Office of Science, Office of Nuclear Physics DOE Contract No. DEAC05-06OR23177, under which Jefferson Science Associates, LLC operates Jefferson Lab.
Construction and operating funding for the experiment was provided through the DOE, the Natural Sciences and Engineering Research Council of Canada (NSERC), and the National Science Foundation (NSF) with university matching contributions from the William \& Mary, Virginia Tech, George Washington University, and Louisiana Tech University. 

\bibliography{qweak_bnssa_proton}

\begin{thebibliography}{39}%
\makeatletter
\providecommand \@ifxundefined [1]{%
 \@ifx{#1\undefined}
}%
\providecommand \@ifnum [1]{%
 \ifnum #1\expandafter \@firstoftwo
 \else \expandafter \@secondoftwo
 \fi
}%
\providecommand \@ifx [1]{%
 \ifx #1\expandafter \@firstoftwo
 \else \expandafter \@secondoftwo
 \fi
}%
\providecommand \natexlab [1]{#1}%
\providecommand \enquote  [1]{``#1''}%
\providecommand \bibnamefont  [1]{#1}%
\providecommand \bibfnamefont [1]{#1}%
\providecommand \citenamefont [1]{#1}%
\providecommand \href@noop [0]{\@secondoftwo}%
\providecommand \href [0]{\begingroup \@sanitize@url \@href}%
\providecommand \@href[1]{\@@startlink{#1}\@@href}%
\providecommand \@@href[1]{\endgroup#1\@@endlink}%
\providecommand \@sanitize@url [0]{\catcode `\\12\catcode `\$12\catcode
  `\&12\catcode `\#12\catcode `\^12\catcode `\_12\catcode `\%12\relax}%
\providecommand \@@startlink[1]{}%
\providecommand \@@endlink[0]{}%
\providecommand \url  [0]{\begingroup\@sanitize@url \@url }%
\providecommand \@url [1]{\endgroup\@href {#1}{\urlprefix }}%
\providecommand \urlprefix  [0]{URL }%
\providecommand \Eprint [0]{\href }%
\providecommand \doibase [0]{http://dx.doi.org/}%
\providecommand \selectlanguage [0]{\@gobble}%
\providecommand \bibinfo  [0]{\@secondoftwo}%
\providecommand \bibfield  [0]{\@secondoftwo}%
\providecommand \translation [1]{[#1]}%
\providecommand \BibitemOpen [0]{}%
\providecommand \bibitemStop [0]{}%
\providecommand \bibitemNoStop [0]{.\EOS\space}%
\providecommand \EOS [0]{\spacefactor3000\relax}%
\providecommand \BibitemShut  [1]{\csname bibitem#1\endcsname}%
\let\auto@bib@innerbib\@empty
\bibitem [{\citenamefont {Afanasev}\ \emph {et~al.}(2017)\citenamefont
  {Afanasev}, \citenamefont {Blunden}, \citenamefont {Hasell},\ and\
  \citenamefont {Raue}}]{Afanasev:2017gsk}%
  \BibitemOpen
  \bibfield  {author} {\bibinfo {author} {\bibfnamefont {A.}~\bibnamefont
  {Afanasev}}, \bibinfo {author} {\bibfnamefont {P.}~\bibnamefont {Blunden}},
  \bibinfo {author} {\bibfnamefont {D.}~\bibnamefont {Hasell}}, \ and\ \bibinfo
  {author} {\bibfnamefont {B.}~\bibnamefont {Raue}},\ }\href {\doibase
  10.1016/j.ppnp.2017.03.004} {\bibfield  {journal} {\bibinfo  {journal}
  {Prog.\ Part.\ Nucl.\ Phys.}\ }\textbf {\bibinfo {volume} {95}},\ \bibinfo
  {pages} {245} (\bibinfo {year} {2017})}\BibitemShut {NoStop}%
\bibitem [{\citenamefont {Barschall}\ and\ \citenamefont
  {Haeberli}(1971)}]{osti_4726823}%
  \BibitemOpen
  \bibinfo {editor} {\bibfnamefont {H.}~\bibnamefont {Barschall}}\ and\
  \bibinfo {editor} {\bibfnamefont {W.}~\bibnamefont {Haeberli}},\ eds.,\
  \href@noop {} {\emph {\bibinfo {title} {Polarization Phenomena in Nuclear
  Reactions, Proc. Third Int. Symposium, Madison, Wisconsin}}}\ (\bibinfo
  {publisher} {U. Wisconson Press},\ \bibinfo {year} {1971})\BibitemShut
  {NoStop}%
\bibitem [{\citenamefont {Gorchtein}\ \emph
  {et~al.}(2004{\natexlab{a}})\citenamefont {Gorchtein}, \citenamefont
  {Guichon},\ and\ \citenamefont {Vanderhaeghen}}]{Gorchtein:2004ac}%
  \BibitemOpen
  \bibfield  {author} {\bibinfo {author} {\bibfnamefont {M.}~\bibnamefont
  {Gorchtein}}, \bibinfo {author} {\bibfnamefont {P.~A.}\ \bibnamefont
  {Guichon}}, \ and\ \bibinfo {author} {\bibfnamefont {M.}~\bibnamefont
  {Vanderhaeghen}},\ }\href {\doibase 10.1016/j.nuclphysa.2004.06.008}
  {\bibfield  {journal} {\bibinfo  {journal} {Nucl.Phys.}\ }\textbf {\bibinfo
  {volume} {A741}},\ \bibinfo {pages} {234} (\bibinfo {year}
  {2004}{\natexlab{a}})}\BibitemShut {NoStop}%
\bibitem [{\citenamefont {Pasquini}\ and\ \citenamefont
  {Vanderhaeghen}(2004)}]{Pasquini:2004pv}%
  \BibitemOpen
  \bibfield  {author} {\bibinfo {author} {\bibfnamefont {B.}~\bibnamefont
  {Pasquini}}\ and\ \bibinfo {author} {\bibfnamefont {M.}~\bibnamefont
  {Vanderhaeghen}},\ }\href {\doibase 10.1103/PhysRevC.70.045206} {\bibfield
  {journal} {\bibinfo  {journal} {Phys.Rev.}\ }\textbf {\bibinfo {volume}
  {C70}},\ \bibinfo {pages} {045206} (\bibinfo {year} {2004})}\BibitemShut
  {NoStop}%
\bibitem [{\citenamefont {Pasquini}()}]{priv:Pasquni2013}%
  \BibitemOpen
  \bibfield  {author} {\bibinfo {author} {\bibfnamefont {B.}~\bibnamefont
  {Pasquini}},\ }\href@noop {} {}\bibinfo {note} {{private
  communication}}\BibitemShut {NoStop}%
\bibitem [{\citenamefont {Gorchtein}(2006)}]{Gorchtein:2005yz}%
  \BibitemOpen
  \bibfield  {author} {\bibinfo {author} {\bibfnamefont {M.}~\bibnamefont
  {Gorchtein}},\ }\href {\doibase 10.1103/PhysRevC.73.055201} {\bibfield
  {journal} {\bibinfo  {journal} {Phys. Rev. C}\ }\textbf {\bibinfo {volume}
  {73}},\ \bibinfo {pages} {055201} (\bibinfo {year} {2006})}\BibitemShut
  {NoStop}%
\bibitem [{\citenamefont {Gorchtein}(2014)}]{Gorchtein:2014hla}%
  \BibitemOpen
  \bibfield  {author} {\bibinfo {author} {\bibfnamefont {M.}~\bibnamefont
  {Gorchtein}},\ }\href {\doibase 10.1103/PhysRevC.90.052201} {\bibfield
  {journal} {\bibinfo  {journal} {Phys. Rev. C}\ }\textbf {\bibinfo {volume}
  {90}},\ \bibinfo {pages} {052201(R)} (\bibinfo {year} {2014})}\BibitemShut
  {NoStop}%
\bibitem [{\citenamefont {Gorchtein}\ \emph
  {et~al.}(2004{\natexlab{b}})\citenamefont {Gorchtein}, \citenamefont
  {Guichon},\ and\ \citenamefont {Vanderhaeghen}}]{GORCHTEIN2004234}%
  \BibitemOpen
  \bibfield  {author} {\bibinfo {author} {\bibfnamefont {M.}~\bibnamefont
  {Gorchtein}}, \bibinfo {author} {\bibfnamefont {P.}~\bibnamefont {Guichon}},
  \ and\ \bibinfo {author} {\bibfnamefont {M.}~\bibnamefont {Vanderhaeghen}},\
  }\href {\doibase https://doi.org/10.1016/j.nuclphysa.2004.06.008} {\bibfield
  {journal} {\bibinfo  {journal} {Nuclear Physics A}\ }\textbf {\bibinfo
  {volume} {741}},\ \bibinfo {pages} {234 } (\bibinfo {year}
  {2004}{\natexlab{b}})}\BibitemShut {NoStop}%
\bibitem [{\citenamefont {Gorchtein}()}]{GorchPC}%
  \BibitemOpen
  \bibfield  {author} {\bibinfo {author} {\bibfnamefont {M.}~\bibnamefont
  {Gorchtein}},\ }\href@noop {} {}\bibinfo {howpublished} {private
  communication}\BibitemShut {NoStop}%
\bibitem [{\citenamefont {Afanasev}\ and\ \citenamefont
  {Merenkov}(2004)}]{Afanasev:2004pu}%
  \BibitemOpen
  \bibfield  {author} {\bibinfo {author} {\bibfnamefont {A.~V.}\ \bibnamefont
  {Afanasev}}\ and\ \bibinfo {author} {\bibfnamefont {N.}~\bibnamefont
  {Merenkov}},\ }\href {\doibase 10.1016/j.physletb.2004.08.023} {\bibfield
  {journal} {\bibinfo  {journal} {Phys.Lett.}\ }\textbf {\bibinfo {volume}
  {B599}},\ \bibinfo {pages} {48} (\bibinfo {year} {2004})}\BibitemShut
  {NoStop}%
\bibitem [{\citenamefont {Afanasev}()}]{AfanasevPC}%
  \BibitemOpen
  \bibfield  {author} {\bibinfo {author} {\bibfnamefont {A.}~\bibnamefont
  {Afanasev}},\ }\href@noop {} {}\bibinfo {howpublished} {private
  communication}\BibitemShut {NoStop}%
\bibitem [{\citenamefont {Mo}\ and\ \citenamefont {Tsai}(1969)}]{Mo:1968cg}%
  \BibitemOpen
  \bibfield  {author} {\bibinfo {author} {\bibfnamefont {L.~W.}\ \bibnamefont
  {Mo}}\ and\ \bibinfo {author} {\bibfnamefont {Y.-S.}\ \bibnamefont {Tsai}},\
  }\href {\doibase 10.1103/RevModPhys.41.205} {\bibfield  {journal} {\bibinfo
  {journal} {Rev.\ Mod.\ Phys.}\ }\textbf {\bibinfo {volume} {41}},\ \bibinfo
  {pages} {205} (\bibinfo {year} {1969})}\BibitemShut {NoStop}%
\bibitem [{\citenamefont {Maximon}\ and\ \citenamefont
  {Tjon}(2000)}]{Maximon:2000hm}%
  \BibitemOpen
  \bibfield  {author} {\bibinfo {author} {\bibfnamefont {L.~C.}\ \bibnamefont
  {Maximon}}\ and\ \bibinfo {author} {\bibfnamefont {J.~A.}\ \bibnamefont
  {Tjon}},\ }\href {\doibase 10.1103/PhysRevC.62.054320} {\bibfield  {journal}
  {\bibinfo  {journal} {Phys.Rev.}\ }\textbf {\bibinfo {volume} {C62}},\
  \bibinfo {pages} {054320} (\bibinfo {year} {2000})}\BibitemShut {NoStop}%
\bibitem [{\citenamefont {Rosenbluth}(1950)}]{Rosenbluth:1950yq}%
  \BibitemOpen
  \bibfield  {author} {\bibinfo {author} {\bibfnamefont {M.}~\bibnamefont
  {Rosenbluth}},\ }\href {\doibase 10.1103/PhysRev.79.615} {\bibfield
  {journal} {\bibinfo  {journal} {Phys.Rev.}\ }\textbf {\bibinfo {volume}
  {79}},\ \bibinfo {pages} {615} (\bibinfo {year} {1950})}\BibitemShut
  {NoStop}%
\bibitem [{\citenamefont {Jones}\ \emph {et~al.}(2000)\citenamefont {Jones}
  \emph {et~al.}}]{Jones:1999rz}%
  \BibitemOpen
  \bibfield  {author} {\bibinfo {author} {\bibfnamefont {M.}~\bibnamefont
  {Jones}} \emph {et~al.} (\bibinfo {collaboration} {Jefferson Lab Hall A
  Collaboration}),\ }\href {\doibase 10.1103/PhysRevLett.84.1398} {\bibfield
  {journal} {\bibinfo  {journal} {Phys.Rev.Lett.}\ }\textbf {\bibinfo {volume}
  {84}},\ \bibinfo {pages} {1398} (\bibinfo {year} {2000})}\BibitemShut
  {NoStop}%
\bibitem [{\citenamefont {Guichon}\ and\ \citenamefont
  {Vanderhaeghen}(2003)}]{Guichon:2003qm}%
  \BibitemOpen
  \bibfield  {author} {\bibinfo {author} {\bibfnamefont {P.~A.~M.}\
  \bibnamefont {Guichon}}\ and\ \bibinfo {author} {\bibfnamefont
  {M.}~\bibnamefont {Vanderhaeghen}},\ }\href {\doibase
  10.1103/PhysRevLett.91.142303} {\bibfield  {journal} {\bibinfo  {journal}
  {Phys.Rev.Lett.}\ }\textbf {\bibinfo {volume} {91}},\ \bibinfo {pages}
  {142303} (\bibinfo {year} {2003})}\BibitemShut {NoStop}%
\bibitem [{\citenamefont {Rachek}\ \emph {et~al.}(2015)\citenamefont {Rachek}
  \emph {et~al.}}]{Rachek:2014fam}%
  \BibitemOpen
  \bibfield  {author} {\bibinfo {author} {\bibfnamefont {I.~A.}\ \bibnamefont
  {Rachek}} \emph {et~al.},\ }\href {\doibase 10.1103/PhysRevLett.114.062005}
  {\bibfield  {journal} {\bibinfo  {journal} {Phys. Rev. Lett.}\ }\textbf
  {\bibinfo {volume} {114}},\ \bibinfo {pages} {062005} (\bibinfo {year}
  {2015})}\BibitemShut {NoStop}%
\bibitem [{\citenamefont {Henderson}\ \emph {et~al.}(2017)\citenamefont
  {Henderson} \emph {et~al.}}]{Henderson:2016dea}%
  \BibitemOpen
  \bibfield  {author} {\bibinfo {author} {\bibfnamefont {B.~S.}\ \bibnamefont
  {Henderson}} \emph {et~al.} (\bibinfo {collaboration} {OLYMPUS}),\ }\href
  {\doibase 10.1103/PhysRevLett.118.092501} {\bibfield  {journal} {\bibinfo
  {journal} {Phys. Rev. Lett.}\ }\textbf {\bibinfo {volume} {118}},\ \bibinfo
  {pages} {092501} (\bibinfo {year} {2017})}\BibitemShut {NoStop}%
\bibitem [{\citenamefont {Adikaram}\ \emph {et~al.}(2015)\citenamefont
  {Adikaram} \emph {et~al.}}]{Adikaram:2014ykv}%
  \BibitemOpen
  \bibfield  {author} {\bibinfo {author} {\bibfnamefont {D.}~\bibnamefont
  {Adikaram}} \emph {et~al.} (\bibinfo {collaboration} {CLAS}),\ }\href
  {\doibase 10.1103/PhysRevLett.114.062003} {\bibfield  {journal} {\bibinfo
  {journal} {Phys. Rev. Lett.}\ }\textbf {\bibinfo {volume} {114}},\ \bibinfo
  {pages} {062003} (\bibinfo {year} {2015})}\BibitemShut {NoStop}%
\bibitem [{\citenamefont {Xiong}\ \emph {et~al.}(2019)\citenamefont {Xiong}
  \emph {et~al.}}]{Xiong:2019umf}%
  \BibitemOpen
  \bibfield  {author} {\bibinfo {author} {\bibfnamefont {W.}~\bibnamefont
  {Xiong}} \emph {et~al.},\ }\href {\doibase 10.1038/s41586-019-1721-2}
  {\bibfield  {journal} {\bibinfo  {journal} {Nature}\ }\textbf {\bibinfo
  {volume} {575}},\ \bibinfo {pages} {147} (\bibinfo {year}
  {2019})}\BibitemShut {NoStop}%
\bibitem [{\citenamefont {Gilman}\ \emph {et~al.}(2013)\citenamefont {Gilman}
  \emph {et~al.}}]{Gilman:2013eiv}%
  \BibitemOpen
  \bibfield  {author} {\bibinfo {author} {\bibfnamefont {R.}~\bibnamefont
  {Gilman}} \emph {et~al.} (\bibinfo {collaboration} {MUSE}),\ }\href@noop {}
  {\  (\bibinfo {year} {2013})},\ \Eprint {http://arxiv.org/abs/1303.2160}
  {arXiv:1303.2160 [nucl-ex]} \BibitemShut {NoStop}%
\bibitem [{\citenamefont {Gilman}\ \emph {et~al.}(2017)\citenamefont {Gilman}
  \emph {et~al.}}]{Gilman:2017hdr}%
  \BibitemOpen
  \bibfield  {author} {\bibinfo {author} {\bibfnamefont {R.}~\bibnamefont
  {Gilman}} \emph {et~al.} (\bibinfo {collaboration} {MUSE}),\ }\href@noop {}
  {\  (\bibinfo {year} {2017})},\ \Eprint {http://arxiv.org/abs/1709.09753}
  {arXiv:1709.09753 [physics.ins-det]} \BibitemShut {NoStop}%
\bibitem [{\citenamefont {De~Rujula}\ \emph {et~al.}(1971)\citenamefont
  {De~Rujula}, \citenamefont {Kaplan},\ and\ \citenamefont
  {De~Rafael}}]{DeRujula:1972te}%
  \BibitemOpen
  \bibfield  {author} {\bibinfo {author} {\bibfnamefont {A.}~\bibnamefont
  {De~Rujula}}, \bibinfo {author} {\bibfnamefont {J.}~\bibnamefont {Kaplan}}, \
  and\ \bibinfo {author} {\bibfnamefont {E.}~\bibnamefont {De~Rafael}},\ }\href
  {\doibase 10.1016/0550-3213(71)90460-3} {\bibfield  {journal} {\bibinfo
  {journal} {Nucl.Phys.}\ }\textbf {\bibinfo {volume} {B35}},\ \bibinfo {pages}
  {365} (\bibinfo {year} {1971})}\BibitemShut {NoStop}%
\bibitem [{\citenamefont {Armstrong}\ \emph {et~al.}(2007)\citenamefont
  {Armstrong} \emph {et~al.}}]{Armstrong:2007vm}%
  \BibitemOpen
  \bibfield  {author} {\bibinfo {author} {\bibfnamefont {D.~S.}\ \bibnamefont
  {Armstrong}} \emph {et~al.} (\bibinfo {collaboration} {G0 Collaboration}),\
  }\href {\doibase 10.1103/PhysRevLett.99.092301} {\bibfield  {journal}
  {\bibinfo  {journal} {Phys.Rev.Lett.}\ }\textbf {\bibinfo {volume} {99}},\
  \bibinfo {pages} {092301} (\bibinfo {year} {2007})}\BibitemShut {NoStop}%
\bibitem [{\citenamefont {Abrahamyan}\ \emph {et~al.}(2012)\citenamefont
  {Abrahamyan} \emph {et~al.}}]{Abrahamyan:2012cg}%
  \BibitemOpen
  \bibfield  {author} {\bibinfo {author} {\bibfnamefont {S.}~\bibnamefont
  {Abrahamyan}} \emph {et~al.} (\bibinfo {collaboration} {HAPPEX, PREX}),\
  }\href {\doibase 10.1103/PhysRevLett.109.192501} {\bibfield  {journal}
  {\bibinfo  {journal} {Phys. Rev. Lett.}\ }\textbf {\bibinfo {volume} {109}},\
  \bibinfo {pages} {192501} (\bibinfo {year} {2012})}\BibitemShut {NoStop}%
\bibitem [{\citenamefont {Maas}\ \emph {et~al.}(2005)\citenamefont {Maas} \emph
  {et~al.}}]{Maas:2004pd}%
  \BibitemOpen
  \bibfield  {author} {\bibinfo {author} {\bibfnamefont {F.~E.}\ \bibnamefont
  {Maas}} \emph {et~al.},\ }\href {\doibase 10.1103/PhysRevLett.94.082001}
  {\bibfield  {journal} {\bibinfo  {journal} {Phys. Rev. Lett.}\ }\textbf
  {\bibinfo {volume} {94}},\ \bibinfo {pages} {082001} (\bibinfo {year}
  {2005})}\BibitemShut {NoStop}%
\bibitem [{\citenamefont {Gou}\ \emph {et~al.}(2020)\citenamefont {Gou} \emph
  {et~al.}}]{Gou:2020viq}%
  \BibitemOpen
  \bibfield  {author} {\bibinfo {author} {\bibfnamefont {B.}~\bibnamefont
  {Gou}} \emph {et~al.},\ }\href {\doibase 10.1103/PhysRevLett.124.122003}
  {\bibfield  {journal} {\bibinfo  {journal} {Phys.\ Rev.\ Lett.}\ }\textbf
  {\bibinfo {volume} {124}},\ \bibinfo {pages} {122003} (\bibinfo {year}
  {2020})}\BibitemShut {NoStop}%
\bibitem [{\citenamefont {Wells}\ \emph {et~al.}(2001)\citenamefont {Wells}
  \emph {et~al.}}]{Wells:2000rx}%
  \BibitemOpen
  \bibfield  {author} {\bibinfo {author} {\bibfnamefont {S.~P.}\ \bibnamefont
  {Wells}} \emph {et~al.} (\bibinfo {collaboration} {SAMPLE}),\ }\href
  {\doibase 10.1103/PhysRevC.63.064001} {\bibfield  {journal} {\bibinfo
  {journal} {Phys. Rev.}\ }\textbf {\bibinfo {volume} {C63}},\ \bibinfo {pages}
  {064001} (\bibinfo {year} {2001})}\BibitemShut {NoStop}%
\bibitem [{\citenamefont {Androi\'c}\ \emph {et~al.}(2011)\citenamefont
  {Androi\'c} \emph {et~al.}}]{Androic:2011rh}%
  \BibitemOpen
  \bibfield  {author} {\bibinfo {author} {\bibfnamefont {D.}~\bibnamefont
  {Androi\'c}} \emph {et~al.} (\bibinfo {collaboration} {G0}),\ }\href
  {\doibase 10.1103/PhysRevLett.107.022501} {\bibfield  {journal} {\bibinfo
  {journal} {Phys. Rev. Lett.}\ }\textbf {\bibinfo {volume} {107}},\ \bibinfo
  {pages} {022501} (\bibinfo {year} {2011})}\BibitemShut {NoStop}%
\bibitem [{\citenamefont {R\'{\i}os}\ \emph {et~al.}(2017)\citenamefont
  {R\'{\i}os} \emph {et~al.}}]{Rios:2017vsw}%
  \BibitemOpen
  \bibfield  {author} {\bibinfo {author} {\bibfnamefont {D.~B.}\ \bibnamefont
  {R\'{\i}os}} \emph {et~al.},\ }\href {\doibase
  10.1103/PhysRevLett.119.012501} {\bibfield  {journal} {\bibinfo  {journal}
  {Phys. Rev. Lett.}\ }\textbf {\bibinfo {volume} {119}},\ \bibinfo {pages}
  {012501} (\bibinfo {year} {2017})}\BibitemShut {NoStop}%
\bibitem [{\citenamefont {Androi\'c}\ \emph {et~al.}(2018)\citenamefont
  {Androi\'c} \emph {et~al.}}]{Qweak.Nature}%
  \BibitemOpen
  \bibfield  {author} {\bibinfo {author} {\bibfnamefont {D.}~\bibnamefont
  {Androi\'c}} \emph {et~al.} (\bibinfo {collaboration} {Q$_{\text{weak}}$
  Collaboration}),\ }\href {\doibase 10.1038/s41586-018-0096-0} {\bibfield
  {journal} {\bibinfo  {journal} {Nature}\ }\textbf {\bibinfo {volume} {557}},\
  \bibinfo {pages} {207} (\bibinfo {year} {2018})}\BibitemShut {NoStop}%
\bibitem [{\citenamefont {Androi\'c}\ \emph {et~al.}(2013)\citenamefont
  {Androi\'c} \emph {et~al.}}]{Androic:2013rhu}%
  \BibitemOpen
  \bibfield  {author} {\bibinfo {author} {\bibfnamefont {D.}~\bibnamefont
  {Androi\'c}} \emph {et~al.} (\bibinfo {collaboration} {$Q_{weak}$
  Collaboration}),\ }\href {\doibase 10.1103/PhysRevLett.111.141803} {\bibfield
   {journal} {\bibinfo  {journal} {Phys. Rev. Lett.}\ }\textbf {\bibinfo
  {volume} {111}},\ \bibinfo {pages} {141803} (\bibinfo {year}
  {2013})}\BibitemShut {NoStop}%
\bibitem [{\citenamefont {Allison}\ \emph {et~al.}(2015)\citenamefont {Allison}
  \emph {et~al.}}]{ALLISON2015105}%
  \BibitemOpen
  \bibfield  {author} {\bibinfo {author} {\bibfnamefont {T.}~\bibnamefont
  {Allison}} \emph {et~al.} (\bibinfo {collaboration} {Q$_{\text{weak}}$
  Collaboration}),\ }\href {\doibase
  https://doi.org/10.1016/j.nima.2015.01.023} {\bibfield  {journal} {\bibinfo
  {journal} {Nucl. Instrum. Methods}\ }\textbf {\bibinfo {volume} {A781}},\
  \bibinfo {pages} {105 } (\bibinfo {year} {2015})}\BibitemShut {NoStop}%
\bibitem [{\citenamefont {James}\ \emph {et~al.}(2011)\citenamefont {James},
  \citenamefont {Adderley}, \citenamefont {Benesch}, \citenamefont {Clark},
  \citenamefont {Hansknecht} \emph {et~al.}}]{Grames:2011}%
  \BibitemOpen
  \bibfield  {author} {\bibinfo {author} {\bibfnamefont {J.}~\bibnamefont
  {James}}, \bibinfo {author} {\bibfnamefont {P.}~\bibnamefont {Adderley}},
  \bibinfo {author} {\bibfnamefont {J.}~\bibnamefont {Benesch}}, \bibinfo
  {author} {\bibfnamefont {J.}~\bibnamefont {Clark}}, \bibinfo {author}
  {\bibfnamefont {J.}~\bibnamefont {Hansknecht}},  \emph {et~al.},\ }in\
  \href@noop {} {\emph {\bibinfo {booktitle} {Proc. of 2011 Particle
  Accelerator Conf.}}}\ (\bibinfo {address} {New York, NY},\ \bibinfo {year}
  {2011})\BibitemShut {NoStop}%
\bibitem [{\citenamefont {Hauger}\ \emph {et~al.}(2001)\citenamefont {Hauger}
  \emph {et~al.}}]{Hauger:1999iv}%
  \BibitemOpen
  \bibfield  {author} {\bibinfo {author} {\bibfnamefont {M.}~\bibnamefont
  {Hauger}} \emph {et~al.},\ }\href {\doibase 10.1016/S0168-9002(01)00197-8}
  {\bibfield  {journal} {\bibinfo  {journal} {Nucl. Instrum. Methods}\ }\textbf
  {\bibinfo {volume} {A462}},\ \bibinfo {pages} {382} (\bibinfo {year}
  {2001})}\BibitemShut {NoStop}%
\bibitem [{\citenamefont {Magee}\ \emph {et~al.}(2017)\citenamefont {Magee}
  \emph {et~al.}}]{Magee:2016xqx}%
  \BibitemOpen
  \bibfield  {author} {\bibinfo {author} {\bibfnamefont {J.~A.}\ \bibnamefont
  {Magee}} \emph {et~al.},\ }\href {\doibase 10.1016/j.physletb.2017.01.026}
  {\bibfield  {journal} {\bibinfo  {journal} {Phys. Lett.}\ }\textbf {\bibinfo
  {volume} {B766}},\ \bibinfo {pages} {339} (\bibinfo {year}
  {2017})}\BibitemShut {NoStop}%
\bibitem [{\citenamefont {Narayan}\ \emph {et~al.}(2016)\citenamefont {Narayan}
  \emph {et~al.}}]{Narayan:2015aua}%
  \BibitemOpen
  \bibfield  {author} {\bibinfo {author} {\bibfnamefont {A.}~\bibnamefont
  {Narayan}} \emph {et~al.},\ }\href {\doibase 10.1103/PhysRevX.6.011013}
  {\bibfield  {journal} {\bibinfo  {journal} {Phys. Rev.}\ }\textbf {\bibinfo
  {volume} {X6}},\ \bibinfo {pages} {011013} (\bibinfo {year}
  {2016})}\BibitemShut {NoStop}%
\bibitem [{\citenamefont {Waidyawansa}(2013)}]{BWaidyawansa_phd}%
  \BibitemOpen
  \bibfield  {author} {\bibinfo {author} {\bibfnamefont {D.~B.~P.}\
  \bibnamefont {Waidyawansa}},\ }\emph {\bibinfo {title} {A 3\% Measurement of
  the Beam Normal Single Spin Asymmetry in Forward Angle Elastic Electron
  Proton Scattering Using the Q$_{weak}$ Setup}},\ \href
  {https://misportal.jlab.org/ul/publications/downloadFile.cfm?pub_id=12540}
  {Ph.D. thesis},\ \bibinfo  {school} {Ohio University} (\bibinfo {year}
  {2013})\BibitemShut {NoStop}%
\bibitem [{\citenamefont {Melnitchouk}\ \emph {et~al.}()\citenamefont
  {Melnitchouk}, \citenamefont {Blunden},\ and\ \citenamefont
  {Sachdeva}}]{priv:WBP}%
  \BibitemOpen
  \bibfield  {author} {\bibinfo {author} {\bibfnamefont {W.}~\bibnamefont
  {Melnitchouk}}, \bibinfo {author} {\bibfnamefont {P.}~\bibnamefont
  {Blunden}}, \ and\ \bibinfo {author} {\bibfnamefont {P.}~\bibnamefont
  {Sachdeva}},\ }\href@noop {} {}\bibinfo {note} {{private
  communication}}\BibitemShut {NoStop}%
\end{thebibliography}%

\end{document}